\documentclass[aps,pra,reprint,groupedaddress,showpacs]{revtex4-1}

\usepackage{amssymb,amsmath}
\usepackage{graphicx}
\usepackage{color}

\begin{document}


\title{Loading and compression of a single two-dimensional Bose gas in an optical accordion}


\author{J.L. Ville$^{1} $}
\author{T. Bienaim\'e$^{2}$}
\author{R. Saint-Jalm$^{1}$}
\author{L. Corman$^{3}$}
\author{M. Aidelsburger$^{1}$}
\author{L. Chomaz$^{4}$}
\author{K. Kleinlein$^{1}$}
\author{D. Perconte$^{5}$}
\author{S. Nascimb\`ene$^{1}$}
\author{J. Dalibard$^{1}$}
\author{J. Beugnon$^{1}$}

\email[]{beugnon@lkb.ens.fr}
\affiliation{$^1$Laboratoire Kastler Brossel, Coll\`ege de France, CNRS, ENS-PSL Research University, UPMC-Sorbonne Universit\'es,
11 place Marcelin-Berthelot, 75005 Paris, France}
\affiliation{$^2$ INO-CNR BEC Center and Dipartimento di Fisica, Universit\`a di Trento, 38123 Povo, Italy}
\affiliation{$^3$Institute for Quantum Electronics, ETH Zurich, 8093 Zurich, Switzerland}
\affiliation{ $^4$ Institut f\"ur Experimentalphysik, Universit\"at Innsbruck, Technikerstra\ss e 25, 6020 Innsbruck, Austria}
\affiliation{ $^5$ Unit\'e Mixte de Physique, CNRS, Thales, Univ. Paris-Sud, Universit\'e Paris-Saclay, 91767 Palaiseau, France}


\date{\today}
\begin{abstract}
The experimental realization of two-dimensional (2D) Bose gases with a tunable interaction strength is an important challenge for the study of ultracold quantum matter. Here we report on the realization of an optical accordion creating a lattice potential with a spacing that can be dynamically tuned between 11 and 2\,$\mu$m. We show that we can load ultracold $^{87}$Rb atoms into a single node of this optical lattice in the large spacing configuration and then  decrease nearly adiabatically the spacing to reach a strong harmonic confinement with frequencies larger than $\omega_z/2\pi=$10\,kHz. Atoms are trapped in an additional flat-bottom in-plane potential that is shaped with a high resolution. By combining these tools we create custom-shaped uniform 2D Bose gases with tunable confinement along the transverse direction and hence with a tunable interaction strength. 
\end{abstract}


\maketitle

\section{Introduction}
Thanks to their high degree of isolation from the environment and the rich toolbox developed from atomic physics, quantum gases are ideal platforms to study strongly correlated systems \cite{Bloch08} or to develop new metrology devices \cite{Cronin09}. A key ingredient is the development of custom-shaped optical potentials allowing one to confine atoms in tunable geometries. Atoms are routinely trapped in low-dimensional setups, optical lattices, or, as recently demonstrated, flat-bottom potentials for three-dimensional (3D)  \cite{Gaunt13} and two-dimensional (2D) \cite{Corman14,Chomaz15} gases. 

Low-dimensional systems are of particular interest for several reasons. The role of thermal and quantum fluctuations is enhanced compared to 3D and leads to rich physics such as the existence of the Berezinskii-Kosterlitz-Thouless superfluid phase in two dimensions \cite{Berezinskii71,Kosterlitz73}. When placed in (artificial) magnetic fields, they can give rise to topological phases of matter similar to those appearing in the quantum Hall effect \cite{Dalibard11}.
From a more technical point of view, 2D systems, now routinely used in ``atomic microscope" experiments \cite{Bakr09,Sherson10}, are well suited to implement high-resolution imaging or trap shaping with a resolution typically better than 1\,$\mu$m, without being limited by a short focal depth or line-of-sight integration. 

In 2D cold atomic clouds the interparticle interactions are characterized by a dimensionless parameter $\tilde g=\sqrt{8\pi}a/\ell_z$ where $a$ is the $s$-wave scattering length and $\ell_z$ is the harmonic oscillator length along the strongly confining direction \cite{Hadzibabic11}. Varying the confinement (hence $\ell_z$) thus opens the possibility of controlling the interaction strength for a fixed value of $a$ and eventually entering the strongly interacting regime for large values of $\tilde g$ \cite{Ha13,Murthy15}.

One of the challenges of realizing 2D systems is to load a large fraction of an initial (3D) Bose-Einstein condensate (BEC) in a single highly anisotropic trap with relatively weak confinement in the $xy$ plane and a strong one along the third ($z$) direction. A possible approach consists of making a single potential minimum, using either phase plates, creating a node of intensity of blue-detuned light \cite{Rath10}, or a tightly focused red-detuned single beam \cite{Clade09}. Another approach consists of making an optical lattice by crossing two interfering beams at a fixed angle. In that case, the lattice spacing and hence the achievable strength of the confinement along the $z$ axis are limited by the requirement of a single node loading \cite{Gelmeke09,Ries15}. Yet another possibility is to use a small spacing lattice, load several planes and then remove atoms in all the planes but one \cite{Stock05}. This procedure may lead to an important atom loss that is detrimental for exploring large systems. Single-plane 2D Bose gases have also been demonstrated in radio-frequency dressed magnetic traps with a moderate transverse confinement \cite{Merloti13} or in more complex  setups involving an evanescent optical field close to a dielectric surface \cite{Gillen09}. 

In this paper we create a single 2D cloud with a large number of atoms and a tunable confinement using a so-called ``optical accordion". It consists of loading atoms in a single node of a large-spacing lattice and then increasing the angle between the two interfering beams to make the confinement stronger. This technique has been demonstrated optically, but not implemented on an atomic cloud, in Refs. \cite{Williams08,Li08} and used to increase the spacing of a lattice trapping ultracold atoms \cite{Alassam10}. Compression of quantum gases has been reported in Ref. \cite{Miranda12} using a different technique involving a reflexion on a dielectric surface. It has also recently been mentioned in Ref. \cite{Mitra16}, without any technical detail or study of the compression process. In this work we demonstrate single-plane loading and a fivefold increase of the trapping frequency of a Bose gas in an optical accordion and study the adiabaticity of the compression stage. With far-detuned light and moderate power we obtain clouds of $10^5$ $^{87}$Rb atoms confined with frequencies $\omega_z/2\pi$  higher than 10\,kHz. We show that this compression can be realized in about 100\,ms with a small amount of additional heating compared to the ideal adiabatic evolution. These experiments are carried out with a flat-bottom in-plane potential.

\section{Accordion optical setup}
\begin{figure}[hbt!!]
\centering
\includegraphics[width=8.2cm]{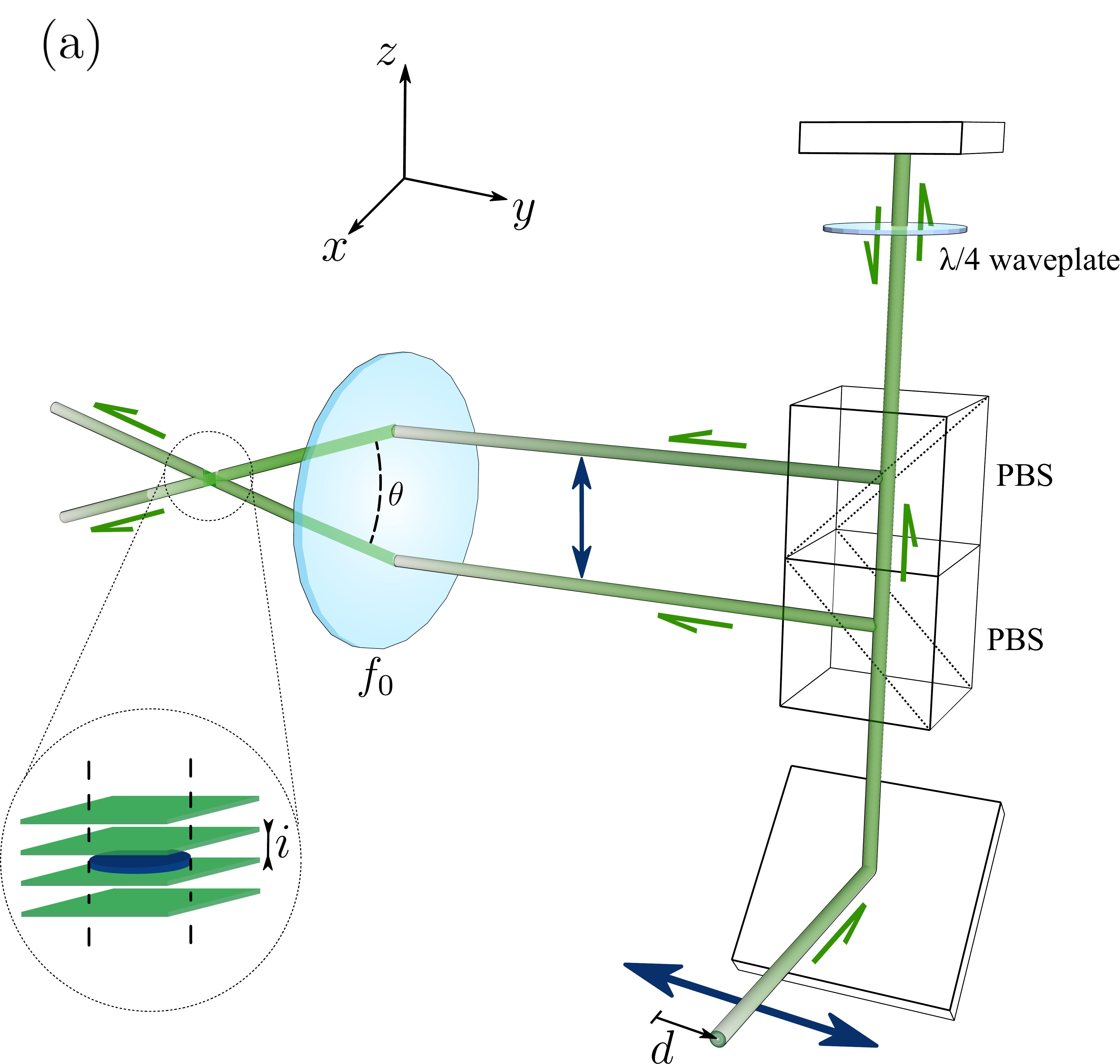}
\includegraphics[width=8.2cm]{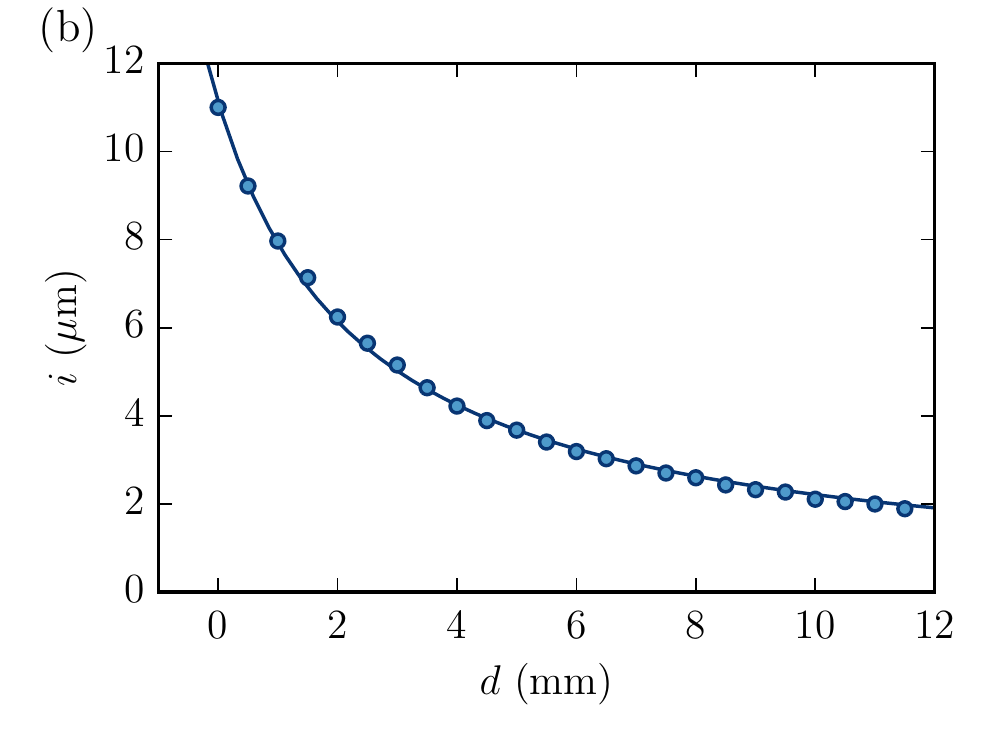}
\caption{(Color online) (a) Sketch of the optical design allowing one to change the angle between the two interfering beams as proposed in \cite{Li08}. The initial beam is moved (bottom arrow) by a distance $d$ with a motorized translation stage (model LS-110 from PI miCos) that changes the distance between the two beams reflected by the polarizing beamsplitters (PBS) of 25\,mm size. These two beams are then focused on the atomic cloud by an aspherical 2-inch-diameter lens of focal length $f_0=100\,$mm. The top beam is reflected on a mirror glued on a piezoelectric stack and goes twice through a quarter-wave plate. (b) Measured lattice spacing $i$ of the vertical lattice at the atom position for different positions $d$ of the translation stage. The data points are fitted by Eq. \eqref{eqinterfringe} with $f$ and $d_0$ as free parameters. We obtain $f=103(1)$\,mm and $d_0=2.46(3)\,$mm. The one-standard-deviation errors obtained from the fit on the measured lattice spacing are smaller than the size of the points.}
\label{figaccordion}
\end{figure}

The design of our accordion lattice is inspired from \cite{Li08} and depicted in Fig.\,\ref{figaccordion}a. A single laser beam of wavelength $\lambda=532\,$nm is split by a pair of polarizing beamsplitters (PBSs) into two parallel beams propagating along the $y$ axis. These two beams cross in the focal plane of a lens and their interference forms a one-dimensional (1D) optical lattice. The position of the incoming beam on the PBSs is moved thanks to a motorized translation stage. This position controls the distance between the two beams reflected by the PBSs, hence the angle between the beams in the focal plane and the fringe spacing. The relative phase between the two beams, which determines the absolute position of the fringes, is controlled by a piezoelectric stack glued on the mirror reflecting the top beam. The two beams are transmitted through a common polarizing beamsplitter cube positioned just before the lens (not shown in Fig.\,\ref{figaccordion}a) to ensure that they have identical polarization. In this work we use an elliptical beam with measured waists at atom positions of $w_x=88(2)\,\mu$m and $w_z=38(6)\,\mu$m, in the horizontal and vertical directions, respectively. The uncertainty corresponds to the standard deviation of the measurement for the different lattice spacings studied here. The choice of these values for the waists results from the compromise between getting the highest intensity with the available power and having a large enough horizontal waist to get a uniform confinement over the sample size (see next section) and a large enough vertical one to ensure a robust overlap between the two beams when changing the lattice spacing as discussed below.

In our setup we change the full angle $\theta$ between the two interfering beams from 3$^\circ$ to 15$^\circ$. The maximum angle is limited by the available numerical aperture on this axis and the minimum angle is constrained by the finite size of the beams, which should not be clipped by the edges of the PBSs. We measure the lattice period $i$ resulting from the interference of the two beams by imaging the intensity pattern in the atom plane on a camera and we obtain the results shown in Fig.\,\ref{figaccordion}b. By translating the initial beam by 11.5\,mm, we vary  $i$ from 11.0(1) to 1.9(1)\,$\mu$m. The data points are fitted by 
\begin{eqnarray}
i=\frac{\lambda}{2} \sqrt{1+[f/(d+d_0)]^2},
\label{eqinterfringe}
\end{eqnarray}
 where $d$ is the displacement of the stage from the position giving our largest lattice period. Here, $d_0$ is an arbitrary offset, and $f$ is the focal length of the lens.

\begin{figure}[hbt!!]
\centering
\includegraphics[width=8.2cm]{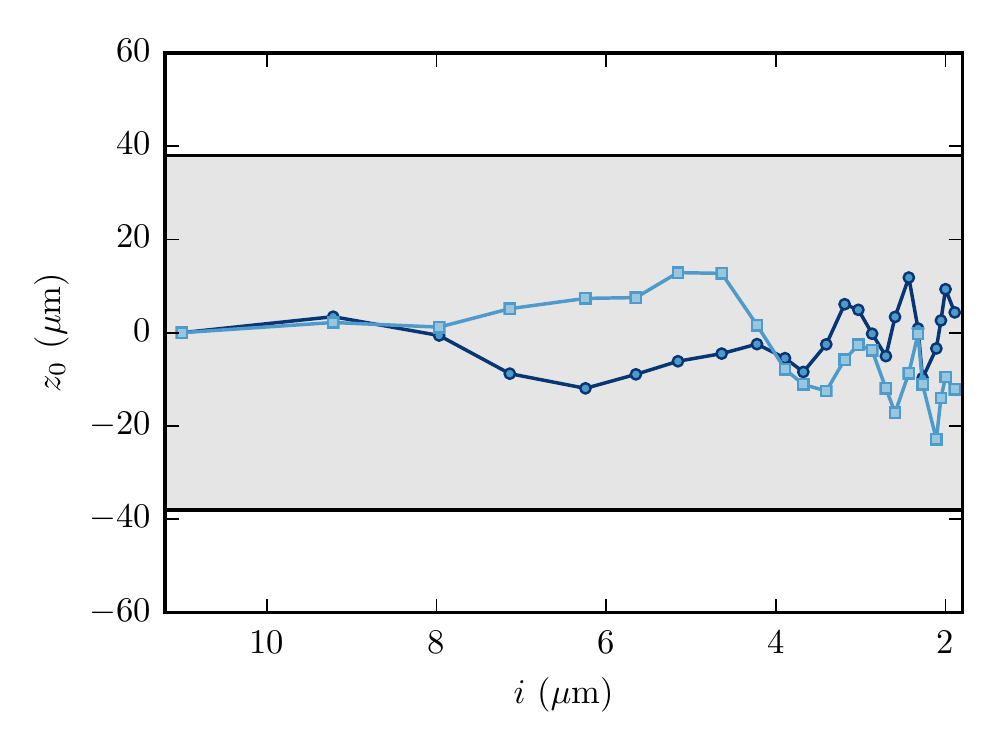}
\caption{(Color online) Variation of the vertical positions $z_0$ with respect to their initial positions of the two interfering beams for different values of the lattice spacing $i$. Squares (resp. circles) correspond to the bottom (top) beam. The shaded area corresponds to $\pm w_z$, where $w_z$ is the averaged measured vertical waist over all the lattice spacings.}
\label{figcenter}
\end{figure}

The main challenge for realizing the accordion lattice is to avoid displacements of the beams in the focal plane when changing their angle. A large displacement of the two beams decreases their overlap and leads to a lower lattice depth, and hence to a reduction of the trapping frequency or even to atom loss. In our setup, the main limitation is the imperfect quality of the lens. For instance, spherical aberrations and surface irregularities induce variations of the beam positions. We have tested standard achromatic doublets and an aspherical lens (Asphericon A50-100) and have found that the displacement is much smaller for the aspherical lens. \footnote{The variations of the beam position and angle induced by the motion of the translation stage were found to be less important than the defects of the tested lenses.}. We show in Fig.\,\ref{figcenter} the positions of the centers of both beams in the $z$ direction. The beams move by typically less than $20\,\mu$m in both directions justifying our choice of $w_z=39$\,$\mu$m. We measure a displacement with a similar amplitude along the horizontal axis. We note that this motion of relatively small amplitude of the beams could induce irregular variations of the trap depth and center that may induce heating when changing the lattice spacing as discussed in Sec. IV.

\section{Making a uniform 2D Bose gas}
\begin{figure}[hbt!!]
\centering
\includegraphics[width=6.2cm]{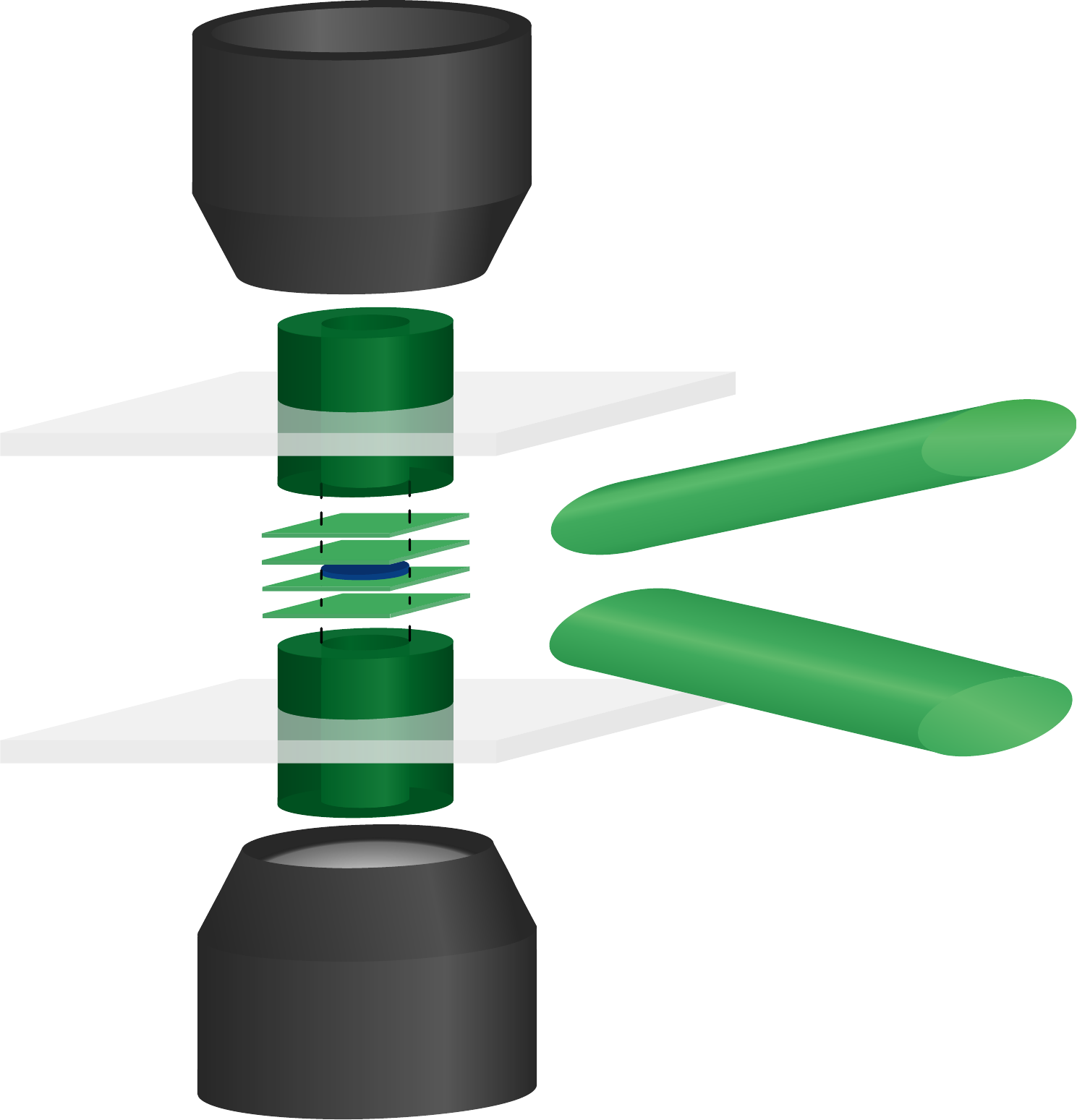}
\caption{(Color online) Sketch of the experimental setup. The vacuum cell, simply depicted here as two horizontal glass plates, is surrounded by a pair of identical microscope objectives with a numerical aperture of 0.45. Atoms (in blue in the center) are trapped in the combination of blue-detuned dipole traps. Confinement along the vertical direction is realized by the interference of two beams at an angle (on the right) that create the accordion lattice. In-plane confinement is ensured by imaging the surface of a DMD on the atomic plane thanks to the top microscope objective. Here we created a disk shaped uniform potential. This trap is loaded from a 3D BEC.}
\label{figsetup}
\end{figure}

We now describe the experimental system and the procedure used to realize 2D uniform gases. A sketch of the setup is shown in Fig.\,\ref{figsetup}. We use two identical microscope objectives (numerical aperture of 0.45) above and below a glass cell. The bottom objective is used for absorption imaging of the cloud on a CCD camera with a typical resolution of 1$\,\mu$m. The top objective allows us to image, with a similar resolution and a magnification of 1/70, a trapping potential programmed on a digital micromirror device (DMD). This spatial light modulator is an array of 1024$\times$784 square mirrors of size 13.8$\,\mu$m. The orientation of each of these mirrors can be chosen between two states. In this work, all the mirrors are set in a state reflecting light towards the atomic cloud except the ones from a central disk-shaped area whose image in the atomic plane has a radius of $20\,\mu$m. The DMD reflects a blue-detuned beam at a wavelength of 532\,nm with a maximum power of about 300\,mW and a waist of 45$\,\mu$m at the atom position. These parameters correspond to a maximum potential height at the edge of the disk of $k_{\rm B}$$\times$4$\,\mu$K. In all the experiments described in the following, atoms are confined in the optical potential created by the combination of this box-potential beam and the accordion beams described in the previous section. The cloud is imaged using standard absorption imaging techniques either along the vertical axis or along the horizontal axis common with the accordion beams.

To load the 2D box potential we first prepare a 3D BEC using standard methods. We start from a 3D magneto-optical trap of $^{87}$Rb atoms which contains $10^9$ atoms. After cooling, compression, and optical pumping into the $F=1$ manifold we load the atoms in  the $F=1$, $m_F=-1$ state in a magnetic quadrupole trap realized by a pair of conical coils along the vertical axis. After compression we proceed to forced evaporative cooling using a radiofrequency field ramp. Afterward we decompress the magnetic trap to load atoms in an optical dipole trap consisting of two beams operating at a wavelength around 1064\,nm and crossing at a right angle in the horizontal plane. Their vertical and horizontal waists are respectively, 30\,$\mu$m and 90\,$\mu$m and the depth potential is calculated to be around 70$\,\mu$K. We then lower the trap depth to realize forced evaporative cooling, and we get almost pure BECs with typically 3$\times 10^5$ atoms.

\begin{figure}[hbt!!]
\centering
\includegraphics[width=7.6cm]{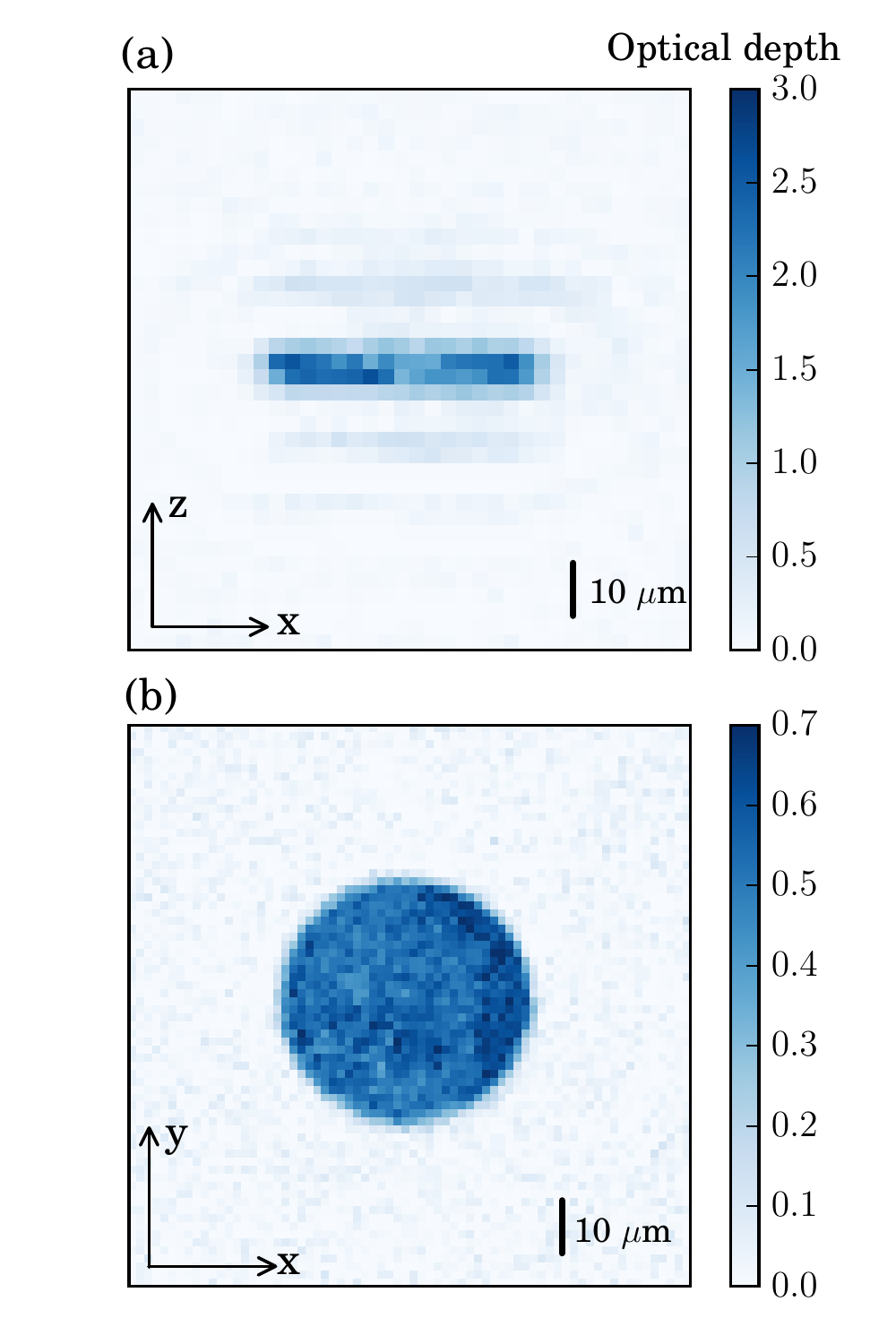}
\caption{(Color online) \textsl{In situ} absorption images of the trapped cloud before compression with $\omega_z/2\pi=2.1$\,kHz and $T=800$\,nK. The cloud diameter is 40$\,\mu$m. To avoid saturation of the absorption signal, we transfer, using a microwave field, only a small fraction of the $10^5$ atoms from the $F=1$ state to the $F=2$ state before imaging atoms in $F=2$. (a) Side view (transferred fraction: 100\% , average of five pictures ). The weak signals above and below the main cloud are fringes due to the propagation of light through our dense sample. We have checked that their position is independent of the lattice spacing of the accordion lattice. (b) Top view (transferred fraction: 2.4\%, average of 35 pictures ).}
\label{figCCD}
\end{figure}

We now detail the loading of the 3D BEC in the box potential. We first ramp the box potential beam to full power in 300\,ms. We then compress the BEC vertically to obtain a robust single-plane loading by increasing the power of one of the red-detuned dipole trap beams back to its maximum initial value in 125\,ms while decreasing the other dipole trap power to zero. We then ramp the power of the accordion beams to their maximal value of 325\,mW per beam in 25\,ms with a maximum spacing of the accordion lattice of 11\,$\mu$m. Finally, we ramp off the crossed dipole trap beams. The global spatial phase of the accordion lattice is adjusted thanks to the piezoelectric stack to get a dark fringe centered on the initial position of the atomic cloud. The optical alignment of the accordion beams is optimized so as to load the atoms in a fringe which is not moving when compressing the accordion lattice. We can then reliably load the atoms in a single plane (see Fig.\,\ref{figCCD}a) \footnote{We compensate the slow drift of the lattice by a feedback on the piezoelectric stack after each experimental sequence.}. Further evaporative cooling can be performed by lowering the power of the box potential beam and/or of the accordion beam to reach the 2D regime for which the thermal energy and the interaction energy are smaller than $\hbar \omega_z$. A typical picture of the cloud taken along the vertical axis is presented in Fig.\,\ref{figCCD}b.

\section{Compression in the accordion}
\begin{figure}[hbt!!]
\centering
\includegraphics[width=8.2cm]{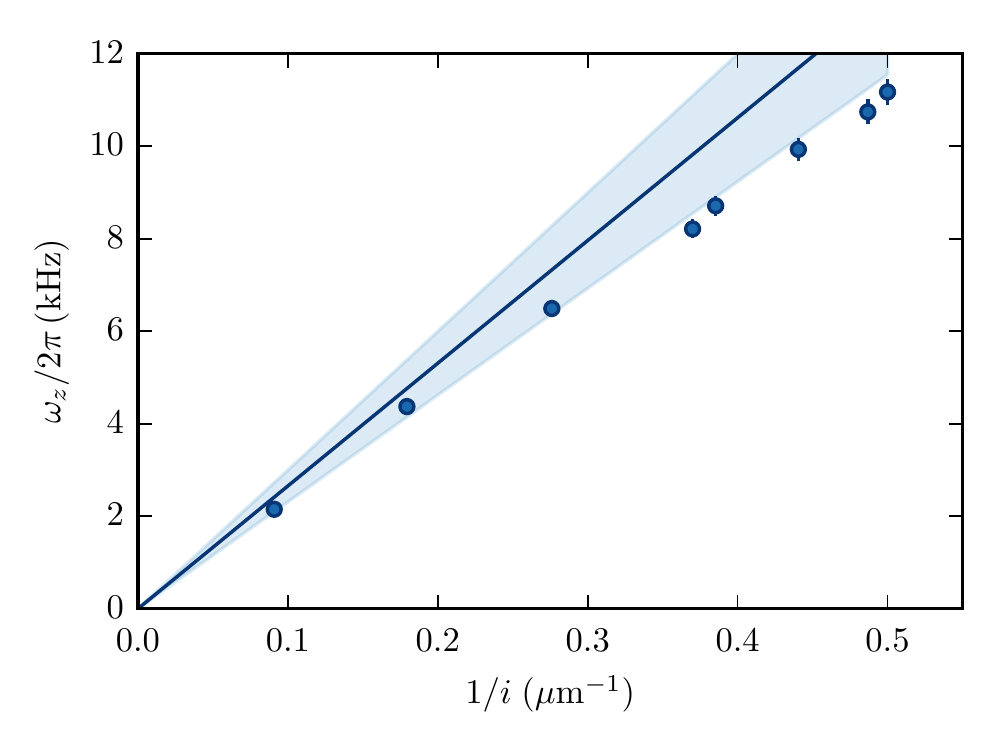}
\caption{(Color online) Measured oscillation frequency along the vertical direction for different lattice spacings. The solid line is the calculated frequency with the independently measured parameters of the beams and the shaded area corresponds to the uncertainty on the calibration of the beam parameters. The error bars represent the standard deviation given by the fit algorithm on the measured frequency and are close to the size of the data points and not visible for the low frequencies.}
\label{figcompfreq}
\end{figure}

The main feature of this setup is the possibility to compress the gas along the $z$ direction once the atoms are loaded in a single node of the lattice. In this section we describe our characterization of the compression process starting from atoms loaded in the largest spacing configuration. First, we measure the oscillation frequency of the cloud in the vertical direction for different lattice spacings at maximum power. This frequency is determined as follows. We excite the center-of-mass motion of the cloud along the $z$ direction by suddenly changing the power in the accordion beams, we let the cloud oscillate, and, finally, we measure the vertical position of the atomic cloud after a short free expansion. The trapping frequency is given by a sinusoidal fit of the data. The results are shown in Fig.\,\ref{figcompfreq}. By compressing the lattice spacing from $11$ to $2\,\mu$m we observe an increase of the oscillation frequency from 2.15(5) to 11.2(3)\,kHz. We also plot in Fig.\,\ref{figcompfreq} the expected frequency calculated with the measured power, waists and lattice spacing. Our measurements are consistently below this calculation. We attribute this effect to the inaccurate calibration of the beam waists and powers and the imperfect overlap of the beams.

\begin{figure}[hbt!!]
\centering
\includegraphics[width=8.2cm]{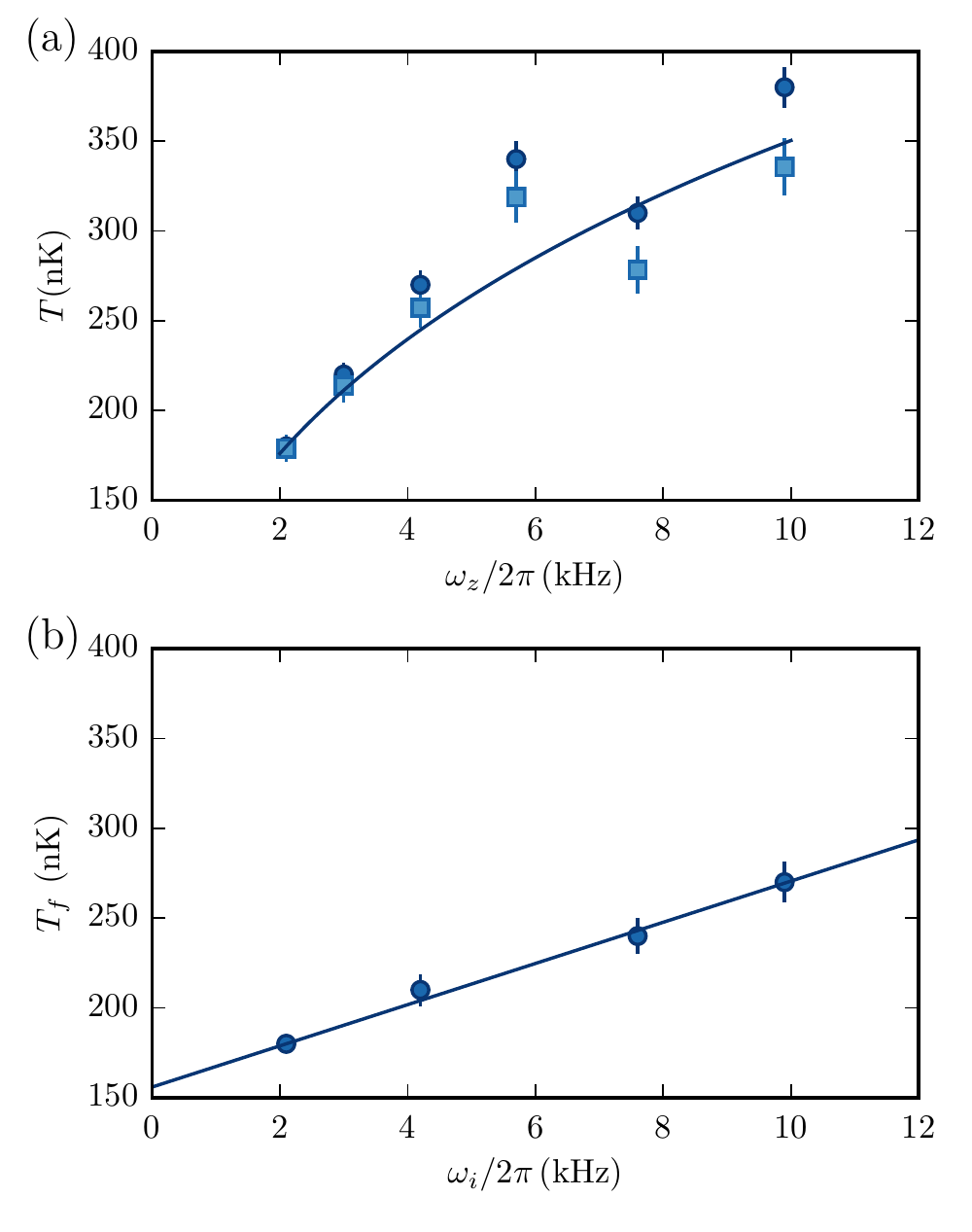}
\caption{(Color online) Compression in the optical accordion. (a) Temperature measured after compression to a final frequency $\omega_z$ (circles) and corrected by the fit of the measured heating displayed in (b) (squares). The solid line is a calculation for an adiabatic compression of an ideal Bose gas with our trap geometry (see Appendix B). (b) Temperature $T_f$ measured after a compression to the intermediate frequency $\omega_i$ and decompression to the initial frequency. The solid line is a linear fit to the data. The measured heating is divided by two before subtraction to the data in (a), considering that the heating for a full cycle is two times larger than the heating for the single compression.}
\label{figcompress}
\end{figure}

We now discuss the effect of compression on the cloud's temperature $T$, which is measured with a method detailed in Appendix A. In order to avoid evaporation of atoms during this compression, we first proceed to a cooling stage. It consists of lowering the power of the in-plane confining laser to evaporate the cloud and then setting it back to its initial value. After this evaporation cooling we typically obtain $N=3\times10^4$ atoms in the large spacing lattice at a temperature of $T_0=180\,$nK. With these parameters, the total 2D phase-space density, defined as $\mathcal{D}=N\lambda_T^2/A$, with $A$ being the disk area and $\lambda_T$ being the thermal de Broglie wavelength, is $\mathcal{D}=4.8$, which corresponds to a non-condensed gas \cite{Chomaz15}. We then compress the cloud  to various final vertical confinements at a constant velocity of the translation stage (90\,mm/s) within 0.13\,s while keeping the overall sequence duration constant. We show in Fig.\,\ref{figcompress}a the measured final temperature (blue circles) for various final trapping frequencies. We observe a significant increase in the cloud's temperature by a factor of about 2 for the largest final frequency.  The atom number is unchanged during this compression, and thus it rules out any evaporation process.

The measured increase of temperature during the compression process could have two origins. (i) It could simply result from the change in density of states in a purely adiabatic process (solid line in Fig.\,\ref{figcompress}a). (ii) There may be an additional heating process due to imperfections in the trap compression as discussed in Sec. II. In order to test the adiabaticity of the process we realize a compression up to a given intermediate frequency $\omega_i$ followed by a decompression to the initial frequency. The measured final temperatures $T_f$ are reported in Fig.\,\ref{figcompress}b. For a purely adiabatic compression-decompression cycle we expect no increase in temperature. We observe a deviation from adiabaticity which can reach 90\,nK for a full compression-decompression sequence or, assuming the same amount of additional heating for compression and decompression, 45\,nK for the compression stage. This heating remains small compared to the 150\,nK increase in temperature expected for a purely adiabatic process as described in the next paragraph. This heating varies approximately linearly with the target frequency $\omega_i$. We have measured a similar heating for lower velocities of the translation stage.

To further explore the origin of the temperature increase observed here, we compare our results to the prediction for adiabatic compression of an ideal Bose gas confined in our trap geometry. The result of this calculation, detailed in Appendix B and applied to the measured initial temperature and frequency, is shown in Fig.\,\ref{figcompress}a as a solid line. We also show the measured temperatures corrected by half the heating measured for the compression-decompression cycle (Fig.\ref{figcompress}b) as squares. They are in good agreement with the calculated temperature. We conclude that the deviation from adiabaticity in our experimental setup leads to an additional heating that remains small compared to the increase in temperature expected in the adiabatic case.

\section{Outlook: an adjustable interaction strength}
We have realized a 2D uniform Bose gas with a tunable confinement. As discussed in the Introduction, in such gases the role of interactions is described, up to logarithmic corrections \cite{Hadzibabic11}, by the dimensionless parameter $\tilde g =\sqrt{8\pi} a/\ell_z$, where $\ell_z=\sqrt{\hbar/(m\omega_z)}$ is the harmonic oscillator groundstate length for a particle of mass $m$ in the harmonic potential of frequency $\omega_z$. Tuning the confinement thus allows one to control the strength of interaction in such systems without tuning the scattering length via a Feshbach resonance \cite{Murthy15} or adding an in-plane lattice potential to control the effective mass of the atoms \cite{Ha13}. In our setup, by varying $\omega_z/2\pi$ between 1 and 11\,kHz by tuning the lattice spacing or the laser power, we can adjust $\tilde g$ between 0.08 and 0.26. Obtaining such comparatively large values of $\tilde g$ is of great interest for realizing strongly correlated states for which the gap between the ground state and the excited states usually scales linearly with $\tilde g$ \cite{Cooper08}. The demonstration of such an optical accordion is thus a significant step in this direction.

Our system is compatible with the realization of flat-bottom potentials with a shape that can be changed, potentially in a dynamic way, thanks to the use of DMDs. Our system is thus an ideal platform to study in- and out-of-equilibrium many-body physics in two-dimensional systems. Another asset of this geometry is the possibility to realize evaporative cooling with this accordion lattice. In the usual evaporation schemes a particle is evaporated when it has a high enough energy and when it reaches a position in the trap where it could be lost (like the edge of the box potential in the work described here). In this situation temperature gradients might be created in the sample. Lowering the depth of the accordion lattice by decreasing its intensity or by adding a magnetic field gradient allows for an evaporation independent of the atom position and could lead to more efficient evaporative cooling. This feature is particularly interesting when studying quench cooling of 2D quantum gases \cite{Chomaz15}.

\vspace{0.2cm}
\begin{acknowledgments}
We thank Z. Hadzibabic for fruitful discussions. This work is supported by DIM NanoK, ANR (ANR-12-BLANAGAFON), and ERC (Synergy UQUAM). L.C. acknowledges the support from DGA. M.A. is  supported  within  the  Marie Curie Project BosQuanTran of the European Commission. J.L.V. and T.B. contributed equally to this work. 
\end{acknowledgments}

\bibliography{accordionbib}

\begin{thebibliography}{28}%
\makeatletter
\providecommand \@ifxundefined [1]{%
 \@ifx{#1\undefined}
}%
\providecommand \@ifnum [1]{%
 \ifnum #1\expandafter \@firstoftwo
 \else \expandafter \@secondoftwo
 \fi
}%
\providecommand \@ifx [1]{%
 \ifx #1\expandafter \@firstoftwo
 \else \expandafter \@secondoftwo
 \fi
}%
\providecommand \natexlab [1]{#1}%
\providecommand \enquote  [1]{``#1''}%
\providecommand \bibnamefont  [1]{#1}%
\providecommand \bibfnamefont [1]{#1}%
\providecommand \citenamefont [1]{#1}%
\providecommand \href@noop [0]{\@secondoftwo}%
\providecommand \href [0]{\begingroup \@sanitize@url \@href}%
\providecommand \@href[1]{\@@startlink{#1}\@@href}%
\providecommand \@@href[1]{\endgroup#1\@@endlink}%
\providecommand \@sanitize@url [0]{\catcode `\\12\catcode `\$12\catcode
  `\&12\catcode `\#12\catcode `\^12\catcode `\_12\catcode `\%12\relax}%
\providecommand \@@startlink[1]{}%
\providecommand \@@endlink[0]{}%
\providecommand \url  [0]{\begingroup\@sanitize@url \@url }%
\providecommand \@url [1]{\endgroup\@href {#1}{\urlprefix }}%
\providecommand \urlprefix  [0]{URL }%
\providecommand \Eprint [0]{\href }%
\providecommand \doibase [0]{http://dx.doi.org/}%
\providecommand \selectlanguage [0]{\@gobble}%
\providecommand \bibinfo  [0]{\@secondoftwo}%
\providecommand \bibfield  [0]{\@secondoftwo}%
\providecommand \translation [1]{[#1]}%
\providecommand \BibitemOpen [0]{}%
\providecommand \bibitemStop [0]{}%
\providecommand \bibitemNoStop [0]{.\EOS\space}%
\providecommand \EOS [0]{\spacefactor3000\relax}%
\providecommand \BibitemShut  [1]{\csname bibitem#1\endcsname}%
\let\auto@bib@innerbib\@empty
\bibitem [{\citenamefont {Bloch}\ \emph {et~al.}(2008)\citenamefont {Bloch},
  \citenamefont {Dalibard},\ and\ \citenamefont {Zwerger}}]{Bloch08}%
  \BibitemOpen
  \bibfield  {author} {\bibinfo {author} {\bibfnamefont {I.}~\bibnamefont
  {Bloch}}, \bibinfo {author} {\bibfnamefont {J.}~\bibnamefont {Dalibard}}, \
  and\ \bibinfo {author} {\bibfnamefont {W.}~\bibnamefont {Zwerger}},\ }\href
  {\doibase 10.1103/RevModPhys.80.885} {\bibfield  {journal} {\bibinfo
  {journal} {Rev. Mod. Phys.}\ }\textbf {\bibinfo {volume} {80}},\ \bibinfo
  {pages} {885} (\bibinfo {year} {2008})}\BibitemShut {NoStop}%
\bibitem [{\citenamefont {Cronin}\ \emph {et~al.}(2009)\citenamefont {Cronin},
  \citenamefont {Schmiedmayer},\ and\ \citenamefont {Pritchard}}]{Cronin09}%
  \BibitemOpen
  \bibfield  {author} {\bibinfo {author} {\bibfnamefont {A.}~\bibnamefont
  {Cronin}}, \bibinfo {author} {\bibfnamefont {J.}~\bibnamefont
  {Schmiedmayer}}, \ and\ \bibinfo {author} {\bibfnamefont {D.}~\bibnamefont
  {Pritchard}},\ }\href {\doibase 10.1103/RevModPhys.81.1051} {\bibfield
  {journal} {\bibinfo  {journal} {Rev. Mod. Phys.}\ }\textbf {\bibinfo {volume}
  {81}},\ \bibinfo {pages} {1051} (\bibinfo {year} {2009})}\BibitemShut
  {NoStop}%
\bibitem [{\citenamefont {Gaunt}\ \emph {et~al.}(2013)\citenamefont {Gaunt},
  \citenamefont {Schmidutz}, \citenamefont {Gotlibovych}, \citenamefont
  {Smith},\ and\ \citenamefont {Hadzibabic}}]{Gaunt13}%
  \BibitemOpen
  \bibfield  {author} {\bibinfo {author} {\bibfnamefont {A.~L.}\ \bibnamefont
  {Gaunt}}, \bibinfo {author} {\bibfnamefont {T.~F.}\ \bibnamefont
  {Schmidutz}}, \bibinfo {author} {\bibfnamefont {I.}~\bibnamefont
  {Gotlibovych}}, \bibinfo {author} {\bibfnamefont {R.~P.}\ \bibnamefont
  {Smith}}, \ and\ \bibinfo {author} {\bibfnamefont {Z.}~\bibnamefont
  {Hadzibabic}},\ }\href@noop {} {\bibfield  {journal} {\bibinfo  {journal}
  {Phys. Rev. Lett.}\ }\textbf {\bibinfo {volume} {110}},\ \bibinfo {pages}
  {200406} (\bibinfo {year} {2013})}\BibitemShut {NoStop}%
\bibitem [{\citenamefont {Corman}\ \emph {et~al.}(2014)\citenamefont {Corman},
  \citenamefont {Chomaz}, \citenamefont {Bienaim\'e}, \citenamefont
  {Desbuquois}, \citenamefont {Weitenberg}, \citenamefont {Nascimb\`ene},
  \citenamefont {Dalibard},\ and\ \citenamefont {Beugnon}}]{Corman14}%
  \BibitemOpen
  \bibfield  {author} {\bibinfo {author} {\bibfnamefont {L.}~\bibnamefont
  {Corman}}, \bibinfo {author} {\bibfnamefont {L.}~\bibnamefont {Chomaz}},
  \bibinfo {author} {\bibfnamefont {T.}~\bibnamefont {Bienaim\'e}}, \bibinfo
  {author} {\bibfnamefont {R.}~\bibnamefont {Desbuquois}}, \bibinfo {author}
  {\bibfnamefont {C.}~\bibnamefont {Weitenberg}}, \bibinfo {author}
  {\bibfnamefont {S.}~\bibnamefont {Nascimb\`ene}}, \bibinfo {author}
  {\bibfnamefont {J.}~\bibnamefont {Dalibard}}, \ and\ \bibinfo {author}
  {\bibfnamefont {J.}~\bibnamefont {Beugnon}},\ }\href {\doibase
  10.1103/PhysRevLett.113.135302} {\bibfield  {journal} {\bibinfo  {journal}
  {Phys. Rev. Lett.}\ }\textbf {\bibinfo {volume} {113}},\ \bibinfo {pages}
  {135302} (\bibinfo {year} {2014})}\BibitemShut {NoStop}%
\bibitem [{\citenamefont {Chomaz}\ \emph {et~al.}(2015)\citenamefont {Chomaz},
  \citenamefont {Corman}, \citenamefont {Bienaim\'e}, \citenamefont
  {Desbuquois}, \citenamefont {Weitenberg}, \citenamefont {Nascimbene},
  \citenamefont {Beugnon},\ and\ \citenamefont {Dalibard}}]{Chomaz15}%
  \BibitemOpen
  \bibfield  {author} {\bibinfo {author} {\bibfnamefont {L.}~\bibnamefont
  {Chomaz}}, \bibinfo {author} {\bibfnamefont {L.}~\bibnamefont {Corman}},
  \bibinfo {author} {\bibfnamefont {T.}~\bibnamefont {Bienaim\'e}}, \bibinfo
  {author} {\bibfnamefont {R.}~\bibnamefont {Desbuquois}}, \bibinfo {author}
  {\bibfnamefont {C.}~\bibnamefont {Weitenberg}}, \bibinfo {author}
  {\bibfnamefont {S.}~\bibnamefont {Nascimbene}}, \bibinfo {author}
  {\bibfnamefont {J.}~\bibnamefont {Beugnon}}, \ and\ \bibinfo {author}
  {\bibfnamefont {J.}~\bibnamefont {Dalibard}},\ }\href@noop {} {\bibfield
  {journal} {\bibinfo  {journal} {{Nat. Commun.}}\ }\textbf {\bibinfo {volume}
  {{6}}},\ \bibinfo {pages} {{6162}} (\bibinfo {year} {{2015}})}\BibitemShut
  {NoStop}%
\bibitem [{\citenamefont {Berezinskii}(1972)}]{Berezinskii71}%
  \BibitemOpen
  \bibfield  {author} {\bibinfo {author} {\bibfnamefont {V.}~\bibnamefont
  {Berezinskii}},\ }\href@noop {} {\bibfield  {journal} {\bibinfo  {journal}
  {{Sov. Phys. JETP}}\ }\textbf {\bibinfo {volume} {{34}}},\ \bibinfo {pages}
  {{610}} (\bibinfo {year} {{1972}})}\BibitemShut {NoStop}%
\bibitem [{\citenamefont {Kostlerlitz}\ and\ \citenamefont
  {Thouless}(1973)}]{Kosterlitz73}%
  \BibitemOpen
  \bibfield  {author} {\bibinfo {author} {\bibfnamefont {J.}~\bibnamefont
  {Kostlerlitz}}\ and\ \bibinfo {author} {\bibfnamefont {D.}~\bibnamefont
  {Thouless}},\ }\href@noop {} {\bibfield  {journal} {\bibinfo  {journal} {{J.
  Phys. C}}\ }\textbf {\bibinfo {volume} {{6}}},\ \bibinfo {pages} {{1181}}
  (\bibinfo {year} {{1973}})}\BibitemShut {NoStop}%
\bibitem [{\citenamefont {Dalibard}\ \emph {et~al.}(2011)\citenamefont
  {Dalibard}, \citenamefont {Gerbier}, \citenamefont
  {Juzeli\ifmmode~\bar{u}\else \={u}\fi{}nas},\ and\ \citenamefont
  {\"Ohberg}}]{Dalibard11}%
  \BibitemOpen
  \bibfield  {author} {\bibinfo {author} {\bibfnamefont {J.}~\bibnamefont
  {Dalibard}}, \bibinfo {author} {\bibfnamefont {F.}~\bibnamefont {Gerbier}},
  \bibinfo {author} {\bibfnamefont {G.}~\bibnamefont
  {Juzeli\ifmmode~\bar{u}\else \={u}\fi{}nas}}, \ and\ \bibinfo {author}
  {\bibfnamefont {P.}~\bibnamefont {\"Ohberg}},\ }\href {\doibase
  10.1103/RevModPhys.83.1523} {\bibfield  {journal} {\bibinfo  {journal} {Rev.
  Mod. Phys.}\ }\textbf {\bibinfo {volume} {83}},\ \bibinfo {pages} {1523}
  (\bibinfo {year} {2011})}\BibitemShut {NoStop}%
\bibitem [{\citenamefont {Bakr}\ \emph {et~al.}(2009)\citenamefont {Bakr},
  \citenamefont {Gillen}, \citenamefont {Peng}, \citenamefont {F{\"o}lling},\
  and\ \citenamefont {Greiner}}]{Bakr09}%
  \BibitemOpen
  \bibfield  {author} {\bibinfo {author} {\bibfnamefont {W.}~\bibnamefont
  {Bakr}}, \bibinfo {author} {\bibfnamefont {J.}~\bibnamefont {Gillen}},
  \bibinfo {author} {\bibfnamefont {A.}~\bibnamefont {Peng}}, \bibinfo {author}
  {\bibfnamefont {S.}~\bibnamefont {F{\"o}lling}}, \ and\ \bibinfo {author}
  {\bibfnamefont {M.}~\bibnamefont {Greiner}},\ }\href@noop {} {\bibfield
  {journal} {\bibinfo  {journal} {Nature}\ }\textbf {\bibinfo {volume} {462}},\
  \bibinfo {pages} {74} (\bibinfo {year} {2009})}\BibitemShut {NoStop}%
\bibitem [{\citenamefont {Sherson}\ \emph {et~al.}(2010)\citenamefont
  {Sherson}, \citenamefont {Weitenberg}, \citenamefont {Endres}, \citenamefont
  {Cheneau}, \citenamefont {Bloch},\ and\ \citenamefont {Kuhr}}]{Sherson10}%
  \BibitemOpen
  \bibfield  {author} {\bibinfo {author} {\bibfnamefont {J.~F.}\ \bibnamefont
  {Sherson}}, \bibinfo {author} {\bibfnamefont {C.}~\bibnamefont {Weitenberg}},
  \bibinfo {author} {\bibfnamefont {M.}~\bibnamefont {Endres}}, \bibinfo
  {author} {\bibfnamefont {M.}~\bibnamefont {Cheneau}}, \bibinfo {author}
  {\bibfnamefont {I.}~\bibnamefont {Bloch}}, \ and\ \bibinfo {author}
  {\bibfnamefont {S.}~\bibnamefont {Kuhr}},\ }\href@noop {} {\bibfield
  {journal} {\bibinfo  {journal} {Nature}\ }\textbf {\bibinfo {volume} {467}},\
  \bibinfo {pages} {68} (\bibinfo {year} {2010})}\BibitemShut {NoStop}%
\bibitem [{\citenamefont {Hadzibabic}\ and\ \citenamefont
  {Dalibard}(2011)}]{Hadzibabic11}%
  \BibitemOpen
  \bibfield  {author} {\bibinfo {author} {\bibfnamefont {Z.}~\bibnamefont
  {Hadzibabic}}\ and\ \bibinfo {author} {\bibfnamefont {J.}~\bibnamefont
  {Dalibard}},\ }\href@noop {} {\bibfield  {journal} {\bibinfo  {journal} {Riv.
  del Nuovo Cimento}\ }\textbf {\bibinfo {volume} {34}},\ \bibinfo {pages}
  {389} (\bibinfo {year} {2011})}\BibitemShut {NoStop}%
\bibitem [{\citenamefont {Ha}\ \emph {et~al.}(2013)\citenamefont {Ha},
  \citenamefont {Hung}, \citenamefont {Zhang}, \citenamefont {Eismann},
  \citenamefont {Tung},\ and\ \citenamefont {Chin}}]{Ha13}%
  \BibitemOpen
  \bibfield  {author} {\bibinfo {author} {\bibfnamefont {L.-C.}\ \bibnamefont
  {Ha}}, \bibinfo {author} {\bibfnamefont {C.-L.}\ \bibnamefont {Hung}},
  \bibinfo {author} {\bibfnamefont {X.}~\bibnamefont {Zhang}}, \bibinfo
  {author} {\bibfnamefont {U.}~\bibnamefont {Eismann}}, \bibinfo {author}
  {\bibfnamefont {S.-K.}\ \bibnamefont {Tung}}, \ and\ \bibinfo {author}
  {\bibfnamefont {C.}~\bibnamefont {Chin}},\ }\href {\doibase
  10.1103/PhysRevLett.110.145302} {\bibfield  {journal} {\bibinfo  {journal}
  {Phys. Rev. Lett.}\ }\textbf {\bibinfo {volume} {110}},\ \bibinfo {pages}
  {145302} (\bibinfo {year} {2013})}\BibitemShut {NoStop}%
\bibitem [{\citenamefont {Murthy}\ \emph {et~al.}(2015)\citenamefont {Murthy},
  \citenamefont {Boettcher}, \citenamefont {Bayha}, \citenamefont {Holzmann},
  \citenamefont {Kedar}, \citenamefont {Neidig}, \citenamefont {Ries},
  \citenamefont {Wenz}, \citenamefont {Z\"urn},\ and\ \citenamefont
  {Jochim}}]{Murthy15}%
  \BibitemOpen
  \bibfield  {author} {\bibinfo {author} {\bibfnamefont {P.~A.}\ \bibnamefont
  {Murthy}}, \bibinfo {author} {\bibfnamefont {I.}~\bibnamefont {Boettcher}},
  \bibinfo {author} {\bibfnamefont {L.}~\bibnamefont {Bayha}}, \bibinfo
  {author} {\bibfnamefont {M.}~\bibnamefont {Holzmann}}, \bibinfo {author}
  {\bibfnamefont {D.}~\bibnamefont {Kedar}}, \bibinfo {author} {\bibfnamefont
  {M.}~\bibnamefont {Neidig}}, \bibinfo {author} {\bibfnamefont {M.~G.}\
  \bibnamefont {Ries}}, \bibinfo {author} {\bibfnamefont {A.~N.}\ \bibnamefont
  {Wenz}}, \bibinfo {author} {\bibfnamefont {G.}~\bibnamefont {Z\"urn}}, \ and\
  \bibinfo {author} {\bibfnamefont {S.}~\bibnamefont {Jochim}},\ }\href
  {\doibase 10.1103/PhysRevLett.115.010401} {\bibfield  {journal} {\bibinfo
  {journal} {Phys. Rev. Lett.}\ }\textbf {\bibinfo {volume} {115}},\ \bibinfo
  {pages} {010401} (\bibinfo {year} {2015})}\BibitemShut {NoStop}%
\bibitem [{\citenamefont {Rath}\ \emph {et~al.}(2010)\citenamefont {Rath},
  \citenamefont {Yefsah}, \citenamefont {G\"unter}, \citenamefont {Cheneau},
  \citenamefont {Desbuquois}, \citenamefont {Holzmann}, \citenamefont
  {Krauth},\ and\ \citenamefont {Dalibard}}]{Rath10}%
  \BibitemOpen
  \bibfield  {author} {\bibinfo {author} {\bibfnamefont {S.~P.}\ \bibnamefont
  {Rath}}, \bibinfo {author} {\bibfnamefont {T.}~\bibnamefont {Yefsah}},
  \bibinfo {author} {\bibfnamefont {K.~J.}\ \bibnamefont {G\"unter}}, \bibinfo
  {author} {\bibfnamefont {M.}~\bibnamefont {Cheneau}}, \bibinfo {author}
  {\bibfnamefont {R.}~\bibnamefont {Desbuquois}}, \bibinfo {author}
  {\bibfnamefont {M.}~\bibnamefont {Holzmann}}, \bibinfo {author}
  {\bibfnamefont {W.}~\bibnamefont {Krauth}}, \ and\ \bibinfo {author}
  {\bibfnamefont {J.}~\bibnamefont {Dalibard}},\ }\href {\doibase
  10.1103/PhysRevA.82.013609} {\bibfield  {journal} {\bibinfo  {journal} {Phys.
  Rev. A}\ }\textbf {\bibinfo {volume} {82}},\ \bibinfo {pages} {013609}
  (\bibinfo {year} {2010})}\BibitemShut {NoStop}%
\bibitem [{\citenamefont {Clad\'e}\ \emph {et~al.}(2009)\citenamefont
  {Clad\'e}, \citenamefont {Ryu}, \citenamefont {Ramanathan}, \citenamefont
  {Helmerson},\ and\ \citenamefont {Phillips}}]{Clade09}%
  \BibitemOpen
  \bibfield  {author} {\bibinfo {author} {\bibfnamefont {P.}~\bibnamefont
  {Clad\'e}}, \bibinfo {author} {\bibfnamefont {C.}~\bibnamefont {Ryu}},
  \bibinfo {author} {\bibfnamefont {A.}~\bibnamefont {Ramanathan}}, \bibinfo
  {author} {\bibfnamefont {K.}~\bibnamefont {Helmerson}}, \ and\ \bibinfo
  {author} {\bibfnamefont {W.~D.}\ \bibnamefont {Phillips}},\ }\href {\doibase
  10.1103/PhysRevLett.102.170401} {\bibfield  {journal} {\bibinfo  {journal}
  {Phys. Rev. Lett.}\ }\textbf {\bibinfo {volume} {102}},\ \bibinfo {pages}
  {170401} (\bibinfo {year} {2009})}\BibitemShut {NoStop}%
\bibitem [{\citenamefont {Gemelke}\ \emph {et~al.}(2009)\citenamefont
  {Gemelke}, \citenamefont {Zhang}, \citenamefont {Hung},\ and\ \citenamefont
  {Chin}}]{Gelmeke09}%
  \BibitemOpen
  \bibfield  {author} {\bibinfo {author} {\bibfnamefont {N.}~\bibnamefont
  {Gemelke}}, \bibinfo {author} {\bibfnamefont {X.}~\bibnamefont {Zhang}},
  \bibinfo {author} {\bibfnamefont {C.-L.}\ \bibnamefont {Hung}}, \ and\
  \bibinfo {author} {\bibfnamefont {C.}~\bibnamefont {Chin}},\ }\href@noop {}
  {\bibfield  {journal} {\bibinfo  {journal} {Nature}\ }\textbf {\bibinfo
  {volume} {460}},\ \bibinfo {pages} {995} (\bibinfo {year}
  {2009})}\BibitemShut {NoStop}%
\bibitem [{\citenamefont {Ries}\ \emph {et~al.}(2015)\citenamefont {Ries},
  \citenamefont {Wenz}, \citenamefont {Z\"urn}, \citenamefont {Bayha},
  \citenamefont {Boettcher}, \citenamefont {Kedar}, \citenamefont {Murthy},
  \citenamefont {Neidig}, \citenamefont {Lompe},\ and\ \citenamefont
  {Jochim}}]{Ries15}%
  \BibitemOpen
  \bibfield  {author} {\bibinfo {author} {\bibfnamefont {M.~G.}\ \bibnamefont
  {Ries}}, \bibinfo {author} {\bibfnamefont {A.~N.}\ \bibnamefont {Wenz}},
  \bibinfo {author} {\bibfnamefont {G.}~\bibnamefont {Z\"urn}}, \bibinfo
  {author} {\bibfnamefont {L.}~\bibnamefont {Bayha}}, \bibinfo {author}
  {\bibfnamefont {I.}~\bibnamefont {Boettcher}}, \bibinfo {author}
  {\bibfnamefont {D.}~\bibnamefont {Kedar}}, \bibinfo {author} {\bibfnamefont
  {P.~A.}\ \bibnamefont {Murthy}}, \bibinfo {author} {\bibfnamefont
  {M.}~\bibnamefont {Neidig}}, \bibinfo {author} {\bibfnamefont
  {T.}~\bibnamefont {Lompe}}, \ and\ \bibinfo {author} {\bibfnamefont
  {S.}~\bibnamefont {Jochim}},\ }\href {\doibase
  10.1103/PhysRevLett.114.230401} {\bibfield  {journal} {\bibinfo  {journal}
  {Phys. Rev. Lett.}\ }\textbf {\bibinfo {volume} {114}},\ \bibinfo {pages}
  {230401} (\bibinfo {year} {2015})}\BibitemShut {NoStop}%
\bibitem [{\citenamefont {Stock}\ \emph {et~al.}(2005)\citenamefont {Stock},
  \citenamefont {Hadzibabic}, \citenamefont {Battelier}, \citenamefont
  {Cheneau},\ and\ \citenamefont {Dalibard}}]{Stock05}%
  \BibitemOpen
  \bibfield  {author} {\bibinfo {author} {\bibfnamefont {S.}~\bibnamefont
  {Stock}}, \bibinfo {author} {\bibfnamefont {Z.}~\bibnamefont {Hadzibabic}},
  \bibinfo {author} {\bibfnamefont {B.}~\bibnamefont {Battelier}}, \bibinfo
  {author} {\bibfnamefont {M.}~\bibnamefont {Cheneau}}, \ and\ \bibinfo
  {author} {\bibfnamefont {J.}~\bibnamefont {Dalibard}},\ }\href {\doibase
  10.1103/PhysRevLett.95.190403} {\bibfield  {journal} {\bibinfo  {journal}
  {Phys. Rev. Lett.}\ }\textbf {\bibinfo {volume} {95}},\ \bibinfo {pages}
  {190403} (\bibinfo {year} {2005})}\BibitemShut {NoStop}%
\bibitem [{\citenamefont {Merloti}\ \emph {et~al.}(2013)\citenamefont
  {Merloti}, \citenamefont {Dubessy}, \citenamefont {Longchambon},
  \citenamefont {Perrin}, \citenamefont {Pottie}, \citenamefont {Lorent},\ and\
  \citenamefont {Perrin}}]{Merloti13}%
  \BibitemOpen
  \bibfield  {author} {\bibinfo {author} {\bibfnamefont {K.}~\bibnamefont
  {Merloti}}, \bibinfo {author} {\bibfnamefont {R.}~\bibnamefont {Dubessy}},
  \bibinfo {author} {\bibfnamefont {L.}~\bibnamefont {Longchambon}}, \bibinfo
  {author} {\bibfnamefont {A.}~\bibnamefont {Perrin}}, \bibinfo {author}
  {\bibfnamefont {P.-E.}\ \bibnamefont {Pottie}}, \bibinfo {author}
  {\bibfnamefont {V.}~\bibnamefont {Lorent}}, \ and\ \bibinfo {author}
  {\bibfnamefont {H.}~\bibnamefont {Perrin}},\ }\href
  {http://stacks.iop.org/1367-2630/15/i=3/a=033007} {\bibfield  {journal}
  {\bibinfo  {journal} {New Journal of Physics}\ }\textbf {\bibinfo {volume}
  {15}},\ \bibinfo {pages} {033007} (\bibinfo {year} {2013})}\BibitemShut
  {NoStop}%
\bibitem [{\citenamefont {Gillen}\ \emph {et~al.}(2009)\citenamefont {Gillen},
  \citenamefont {Bakr}, \citenamefont {Peng}, \citenamefont {Unterwaditzer},
  \citenamefont {F\"olling},\ and\ \citenamefont {Greiner}}]{Gillen09}%
  \BibitemOpen
  \bibfield  {author} {\bibinfo {author} {\bibfnamefont {J.~I.}\ \bibnamefont
  {Gillen}}, \bibinfo {author} {\bibfnamefont {W.~S.}\ \bibnamefont {Bakr}},
  \bibinfo {author} {\bibfnamefont {A.}~\bibnamefont {Peng}}, \bibinfo {author}
  {\bibfnamefont {P.}~\bibnamefont {Unterwaditzer}}, \bibinfo {author}
  {\bibfnamefont {S.}~\bibnamefont {F\"olling}}, \ and\ \bibinfo {author}
  {\bibfnamefont {M.}~\bibnamefont {Greiner}},\ }\href {\doibase
  10.1103/PhysRevA.80.021602} {\bibfield  {journal} {\bibinfo  {journal} {Phys.
  Rev. A}\ }\textbf {\bibinfo {volume} {80}},\ \bibinfo {pages} {021602}
  (\bibinfo {year} {2009})}\BibitemShut {NoStop}%
\bibitem [{\citenamefont {Williams}\ \emph {et~al.}(2008)\citenamefont
  {Williams}, \citenamefont {Pillet}, \citenamefont {Al-Assam}, \citenamefont
  {Fletcher}, \citenamefont {Shotter},\ and\ \citenamefont
  {Foot}}]{Williams08}%
  \BibitemOpen
  \bibfield  {author} {\bibinfo {author} {\bibfnamefont {R.~A.}\ \bibnamefont
  {Williams}}, \bibinfo {author} {\bibfnamefont {J.~D.}\ \bibnamefont
  {Pillet}}, \bibinfo {author} {\bibfnamefont {S.}~\bibnamefont {Al-Assam}},
  \bibinfo {author} {\bibfnamefont {B.}~\bibnamefont {Fletcher}}, \bibinfo
  {author} {\bibfnamefont {M.}~\bibnamefont {Shotter}}, \ and\ \bibinfo
  {author} {\bibfnamefont {C.~J.}\ \bibnamefont {Foot}},\ }\href {\doibase
  10.1364/OE.16.016977} {\bibfield  {journal} {\bibinfo  {journal} {Opt.
  Express}\ }\textbf {\bibinfo {volume} {16}},\ \bibinfo {pages} {16977}
  (\bibinfo {year} {2008})}\BibitemShut {NoStop}%
\bibitem [{\citenamefont {Li}\ \emph {et~al.}(2008)\citenamefont {Li},
  \citenamefont {Kelkar}, \citenamefont {Medellin},\ and\ \citenamefont
  {Raizen}}]{Li08}%
  \BibitemOpen
  \bibfield  {author} {\bibinfo {author} {\bibfnamefont {T.}~\bibnamefont
  {Li}}, \bibinfo {author} {\bibfnamefont {H.}~\bibnamefont {Kelkar}}, \bibinfo
  {author} {\bibfnamefont {D.}~\bibnamefont {Medellin}}, \ and\ \bibinfo
  {author} {\bibfnamefont {M.}~\bibnamefont {Raizen}},\ }\href@noop {}
  {\bibfield  {journal} {\bibinfo  {journal} {Opt. Express}\ }\textbf {\bibinfo
  {volume} {16}},\ \bibinfo {pages} {5465} (\bibinfo {year}
  {2008})}\BibitemShut {NoStop}%
\bibitem [{\citenamefont {Al-Assam}\ \emph {et~al.}(2010)\citenamefont
  {Al-Assam}, \citenamefont {Williams},\ and\ \citenamefont
  {Foot}}]{Alassam10}%
  \BibitemOpen
  \bibfield  {author} {\bibinfo {author} {\bibfnamefont {S.}~\bibnamefont
  {Al-Assam}}, \bibinfo {author} {\bibfnamefont {R.~A.}\ \bibnamefont
  {Williams}}, \ and\ \bibinfo {author} {\bibfnamefont {C.~J.}\ \bibnamefont
  {Foot}},\ }\href {\doibase 10.1103/PhysRevA.82.021604} {\bibfield  {journal}
  {\bibinfo  {journal} {Phys. Rev. A}\ }\textbf {\bibinfo {volume} {82}},\
  \bibinfo {pages} {021604} (\bibinfo {year} {2010})}\BibitemShut {NoStop}%
\bibitem [{\citenamefont {Miranda}\ \emph {et~al.}(2012)\citenamefont
  {Miranda}, \citenamefont {Nakamoto}, \citenamefont {Okuyama}, \citenamefont
  {Noguchi}, \citenamefont {Ueda},\ and\ \citenamefont {Kozuma}}]{Miranda12}%
  \BibitemOpen
  \bibfield  {author} {\bibinfo {author} {\bibfnamefont {M.}~\bibnamefont
  {Miranda}}, \bibinfo {author} {\bibfnamefont {A.}~\bibnamefont {Nakamoto}},
  \bibinfo {author} {\bibfnamefont {Y.}~\bibnamefont {Okuyama}}, \bibinfo
  {author} {\bibfnamefont {A.}~\bibnamefont {Noguchi}}, \bibinfo {author}
  {\bibfnamefont {M.}~\bibnamefont {Ueda}}, \ and\ \bibinfo {author}
  {\bibfnamefont {M.}~\bibnamefont {Kozuma}},\ }\href {\doibase
  10.1103/PhysRevA.86.063615} {\bibfield  {journal} {\bibinfo  {journal} {Phys.
  Rev. A}\ }\textbf {\bibinfo {volume} {86}},\ \bibinfo {pages} {063615}
  (\bibinfo {year} {2012})}\BibitemShut {NoStop}%
\bibitem [{\citenamefont {{Mitra}}\ \emph {et~al.}(2016)\citenamefont
  {{Mitra}}, \citenamefont {{Brown}}, \citenamefont {{Schau{\ss}}},
  \citenamefont {{Kondov}},\ and\ \citenamefont {{Bakr}}}]{Mitra16}%
  \BibitemOpen
  \bibfield  {author} {\bibinfo {author} {\bibfnamefont {D.}~\bibnamefont
  {{Mitra}}}, \bibinfo {author} {\bibfnamefont {P.~T.}\ \bibnamefont
  {{Brown}}}, \bibinfo {author} {\bibfnamefont {P.}~\bibnamefont
  {{Schau{\ss}}}}, \bibinfo {author} {\bibfnamefont {S.~S.}\ \bibnamefont
  {{Kondov}}}, \ and\ \bibinfo {author} {\bibfnamefont {W.~S.}\ \bibnamefont
  {{Bakr}}},\ }\href {\doibase 10.1103/PhysRevLett.117.093601} {\bibfield
  {journal} {\bibinfo  {journal} {Phys. Rev. Lett.}\ }\textbf {\bibinfo
  {volume} {117}},\ \bibinfo {pages} {093601} (\bibinfo {year}
  {2016})}\BibitemShut {NoStop}%
\bibitem [{Note1()}]{Note1}%
  \BibitemOpen
  \bibinfo {note} {The variations of the beam position and angle induced by the
  motion of the translation stage were found to be less important than the
  defects of the tested lenses.}\BibitemShut {Stop}%
\bibitem [{Note2()}]{Note2}%
  \BibitemOpen
  \bibinfo {note} {We compensate the slow drift of the lattice by a feedback on
  the piezoelectric stack after each experimental sequence.}\BibitemShut
  {Stop}%
\bibitem [{\citenamefont {Cooper}(2008)}]{Cooper08}%
  \BibitemOpen
  \bibfield  {author} {\bibinfo {author} {\bibfnamefont {N.}~\bibnamefont
  {Cooper}},\ }\href@noop {} {\bibfield  {journal} {\bibinfo  {journal}
  {Advances in Physics}\ }\textbf {\bibinfo {volume} {57}},\ \bibinfo {pages}
  {539} (\bibinfo {year} {2008})}\BibitemShut {NoStop}%
\end{thebibliography}%


\appendix

\section{Temperature measurement}
The temperatures reported in the main text were measured using the following procedure. Immediately after loading the atoms in the optical box potential, we send a short pulse of a microwave field that transfers a small fraction of about 10\,\% of the atoms from the $|1\rangle=|F=1,m=-1\rangle$ state to the $|2\rangle=|F=2,m=-2\rangle$ state. We then proceed to forced evaporation by lowering the power of the box potential beam and realize the experimental sequences discussed in the main text. We assume that the atoms in state $|2\rangle$ thermalize with the main cloud of atoms in $|1\rangle$. By choosing the fraction of atoms in state $|2\rangle$ to be small enough we always prevent the formation of a Bose-Einstein condensate in this state. To extract the temperature of the sample we release the atoms from the trap and image the density distribution of atoms in state $|2\rangle$ integrated along the vertical direction and after a time-of-flight of 8.7\,ms. For each point we average typically 10 images with the same experimental conditions.

We compare the radially averaged profile of these distributions to a numerically computed profile considering an ideal gas with an initial velocity distribution given by the Bose distribution, and an initial uniform position in the box potential, and assuming an expansion without any interparticle interaction. The theoretical profile has two free parameters, the temperature $T$ and the fugacity $z$, which we optimize to obtain the best fit to the experimental data points. With our signal-to-noise ratio, there is a continuous set of $(z,T)$ that fits almost equally well a given experimental profile making a robust estimate of the temperature difficult. We circumvent this issue by using the independently measured atom number as an additional input parameter to compute $z(T)$, leaving $T$ as the only free parameter.

From the distribution of temperature measurements for a fixed experimental sequence, we estimate that the one-standard-deviation statistical error bars on the temperature measurement are around $\pm3\,\%$ of the measured temperature. The main source of uncertainty is given by the uncertainty on the atom number that we use to estimate the temperature. In our range of parameters, the estimated uncertainty of 25\% in the atom number calibration leads to an uncertainty of about 15\,\% in the temperature. In the main text, we display error bars corresponding to only the $\pm$3\,\% statistical uncertainty.

\section{Adiabatic compression}
\begin{figure}[hbt!!]
\centering
\includegraphics[width=8.2cm]{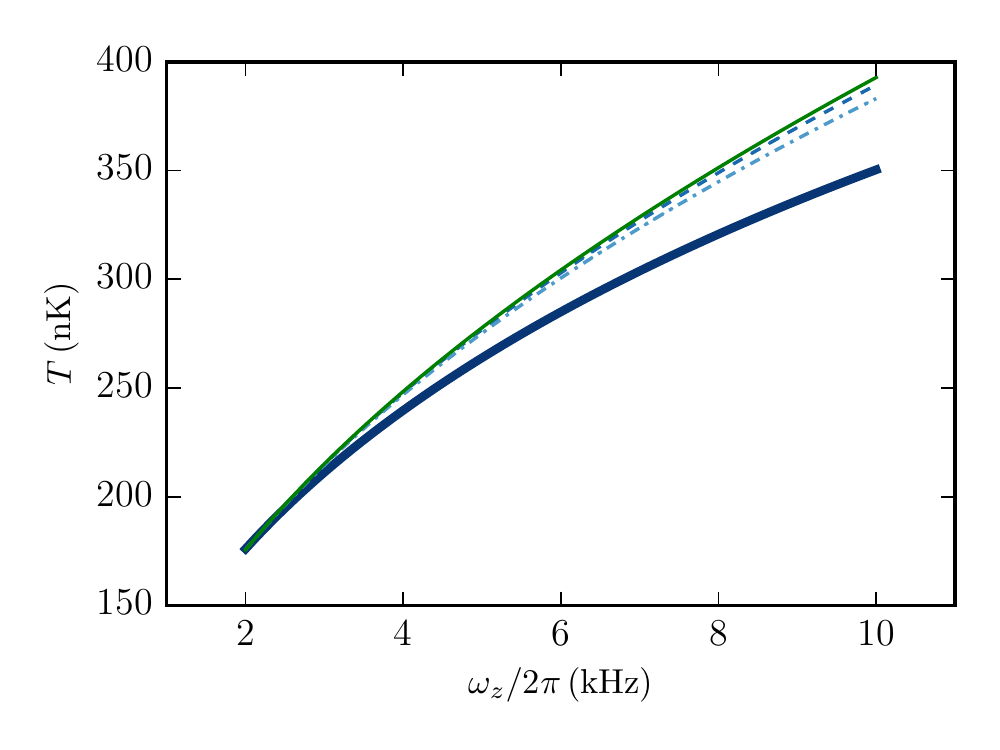}
\caption{(Color online) Adiabatic compression. We show, for an initial temperature of 180\,nK and an initial frequency of 2.1\,kHz, the temperature increase during compression calculated numerically for different models. The thick solid line corresponds to the bosonic case. The green solid line is given by an analytical result obtained in the classical case with a weak confinement along $z$ and scales as $\omega_z^{1/2}$. The dashed line and the dot-dashed line are associated with the fermionic and the Maxwell-Boltzmann statistics cases, respectively.}
\label{figadiabatictheo}
\end{figure}

We consider a gas of non-condensed bosons of mass $m$ confined in the $xy$ plane in a box potential of surface $A$ and along the vertical direction in a harmonic potential of frequency $\omega_z$. We set $\rho_0=mA/(2\pi \hbar^2)$ for the in-plane density of states and $z_j=z \exp(-j \beta \hbar \omega_z)$ where $z$ is the gas fugacity, $\beta=1/k_B T$, and $j$ is the integer labeling the $j$th state of the vertical harmonic oscillator. The average occupation number $\bar{n}_{j,k}$ of a given energy state with an in-plane wavenumber $k$ is
\begin{eqnarray}
\bar{n}_{j,k}=\left( z^{-1} \exp\left[\beta\left(j\hbar \omega_z+\frac{\hbar^2 k^2}{2 m}\right)-1\right]\right)^{-1}.
\end{eqnarray}
Introducing the polylogarithm function $g_\alpha(z)=\sum_{k=1}^\infty z^k/k^\alpha$, we compute $N_j$, $J_j$ and $S_J$ which are respectively the atom number, the grand-canonical potential and the entropy of state $j$.
\begin{eqnarray}
&&N_j=  \rho_0 k_B T g_1(z_j)  \nonumber\\
&&J_j= -\rho_0 (k_B T)^2 g_2(z_j) \nonumber\\
&&S_j=-\frac{\partial J_j}{\partial T}\bigg|_{\mu,\omega_z}=\rho_0 k_B T (2 g_2(z_j)-g_1(z_j) \ln(z_j))\nonumber \\
\end{eqnarray}
We compute the temperature evolution for an adiabatic compression by evaluating, for each value of the final compression frequency $\omega_z$, the temperature and the fugacity keeping $S=\sum_j S_j$ and $N=\sum_j N_j$ constant. The result of this calculation is shown in the main text in Fig.\,\ref{figcompress} and is reproduced in Fig.\,\ref{figadiabatictheo}.

The previous calculation can be straightforwardly extended to fermionic statistics and to the classical Maxwell-Boltzmann statistics by replacing $g_\alpha(z)$ by $f_\alpha(z)=-g_\alpha(-z)$ and by $z$, respectively. The results for these cases are also represented in Fig.\,\ref{figadiabatictheo} respectively as a dotted  line and  a dot-dashed line, respectively, and show that, in all cases, the increase in temperature during adiabatic compression is larger than for the bosonic case. Indeed, Bose statistics leads to a larger population of the low-lying states of the vertical harmonic oscillator than the classical distribution and thus to a smaller increase in temperature when increasing the confinement frequency. 

Finally, we also plot in Fig.\,\ref{figadiabatictheo} an analytical result obtained for the classical Maxwell-Boltzmann statistics but assuming a weak confinement along the vertical direction ($\beta\hbar \omega_z\ll1)$. In this case the 3D density of states is given by $\rho(\varepsilon)=\rho_0/(\hbar\omega_z)\, \varepsilon$, and the entropy reads:
\begin{eqnarray}
S/(N k_B)=3+\ln\left[\rho_0 (k_B T)^2/(N \hbar \omega_z)\right].
\end{eqnarray}
An adiabatic compression thus leads to an increase in temperature as $T \propto \sqrt{\omega_z}$, which corresponds to the green solid line shown in Fig.\,\ref{figadiabatictheo} and which is very close to the numerical calculation for the Maxwell-Boltzmann statistics. We note that in the experiments presented here the fugacity is close to 1 and the Maxwell-Boltzmann approximation is clearly not valid.

\end{document}